\documentclass[natbib]{emulateapj}

\input epsf

\def\s2n{S^{\prime}/N}

\begin{document}
\title{The Power Spectrum of Supersonic Turbulence in Perseus}
\author{Paolo Padoan\altaffilmark{1}, Mika Juvela\altaffilmark{2},
Alexei Kritsuk\altaffilmark{1} and Michael L. Norman\altaffilmark{1}}
\altaffiltext{1}{Department of Physics, University of California, San Diego, 
CASS/UCSD 0424, 9500 Gilman Drive, La Jolla, CA 92093-0424; ppadoan@ucsd.edu}
\altaffiltext{2}{Department of Astronomy, University of Helsinki,
T\"ahtitorninm\"aki, P.O.Box 14,FI-00014 University of Helsinki, Finland}

\begin{abstract}

We test a method of estimating the power spectrum of turbulence in molecular 
clouds based on the comparison of power spectra of integrated intensity maps 
and single-velocity-channel maps, suggested by Lazarian and Pogosyan.
We use synthetic $^{13}$CO data from non-LTE radiative transfer calculations 
based on density and velocity fields of a simulation of supersonic hydrodynamic 
turbulence. We find that the method yields the correct power spectrum with 
good accuracy. We then apply the method to the Five College Radio Astronomy 
Observatory $^{13}$CO map of the Perseus region, from the COMPLETE website. 
We find a power law power spectrum with slope $\beta=1.81\pm0.10$. 
The values of $\beta$ as a function of velocity resolution are also 
confirmed using the lower resolution map of the same region obtained with the 
AT\&T Bell Laboratories antenna. Because of its small uncertainty, this result 
provides a useful constraint for numerical codes used to simulate molecular cloud 
turbulence. 

\end{abstract}

\keywords{
ISM: clouds -- ISM: kinematics and dynamics -- ISM: structure 
}

\section{Introduction}

The large scale dynamics of interstellar clouds is characterized by
random supersonic motions with a very large Reynolds number, meaning
that inertia is much larger than viscous forces. This interstellar 
turbulence is a dominant transport mechanism in many astrophysical 
processes and plays an important role in the fragmentation of 
star-forming clouds. Random supersonic flows in molecular clouds 
result in a complex network of highly radiative shocks causing very 
large density contrasts and shaping the clouds into the observed 
self-similar filamentary structure (Nordlund \& Padoan 2003).
 
Our understanding of turbulence is limited by the lack of general 
analytical solutions of the Navier-Stokes equation. Statistical 
properties of turbulence, essential to modeling many 
astrophysical processes, are therefore derived almost entirely 
from numerical simulations. In the case of incompressible hydrodynamic
turbulence, numerical simulations can be compared with laboratory 
experiments. This is not possible for the isothermal, supersonic, 
magneto-hydrodynamic (MHD) turbulence regime of molecular 
clouds. The best approach to validate numerical models is 
in this case a statistical comparison of observational data with 
{\it synthetic data} from numerical simulations.

Padoan et al. (2004) found that the exponents of the velocity structure 
functions of compressible and super-Alfv\'{e}nic turbulence follow a 
generalized She-L\'{e}v\^{e}que scaling (Boldyrev 2002), depending only 
on the rms Mach number of the flow. Based on that scaling formula, if 
one of the exponents is known (for example the second order that 
corresponds to the velocity power spectrum) all the others can be 
derived, up to some order. While the scaling formula is well constrained 
numerically, its normalization, given by the actual value of one of the 
exponents, is harder to measure and may depend on the numerical method used
to simulate the turbulence. 

Numerical simulations of supersonic hydrodynamic (HD) and magneto-hydrodynamic 
(MHD) turbulence have yielded a range of values of the slope of the velocity
power spectrum, from the Kolmogorov value of $\beta=5/3$, to the Burgers 
value of $\beta=2$ (Frisch \& Bec 2001), and beyond. The main problems with 
the numerical estimate of the power spectrum are i) The limited extent in wavenumbers 
of the inertial range of turbulence, or even its complete absence in the case
of low resolution (or highly dissipative) simulations; ii) The emergence 
of the bottleneck effect (e.g. Falkovich 1994; Dobler et al. 2003; 
Haugen \& Brandenburg 2004) as soon as the numerical resolution is large 
enough to generate an inertial range; iii) The dependence of the 
power spectrum on the numerical schemes used to stabilize the shocks;
iv) The dependence of the numerical resolution necessary for convergence 
on the numerical method. Given these difficulties, reliable measurements from 
interstellar clouds provide useful constraints to validate numerical models and 
to improve our knowledge of supersonic turbulence. 

Lazarian \& Pogosyan (2000) have demonstrated that the exponent of the 
velocity power spectrum can in principle be derived from spectral maps 
of emission lines by comparing the power spectrum of integrated intensity
with the power spectrum of single-velocity-channel intensity. Their method was
tested with numerical simulations of turbulent flows, where the velocity
and density fields had to be modified to generate power law power spectra,
due to the limited numerical resolution (Esquivel et al. 2003). In this 
Letter we present a new test of their method. Thanks to the high numerical 
resolution of our simulation ($1,024^3$ computational zones), our density 
and velocity power spectra exhibit extended power laws, and no modifications 
to the original turbulent fields are required. Furthermore, the method is 
tested by computing synthetic CO emission lines with a non-LTE radiative 
transfer code. We find the method allows to retrieve the power
law exponent of the velocity power spectrum with good accuracy, using the 
J=1-0 $^{13}$CO line. We then apply the method to the Five College 
\begin{figure}[ht]
\centering
\epsfxsize=9cm \epsfbox{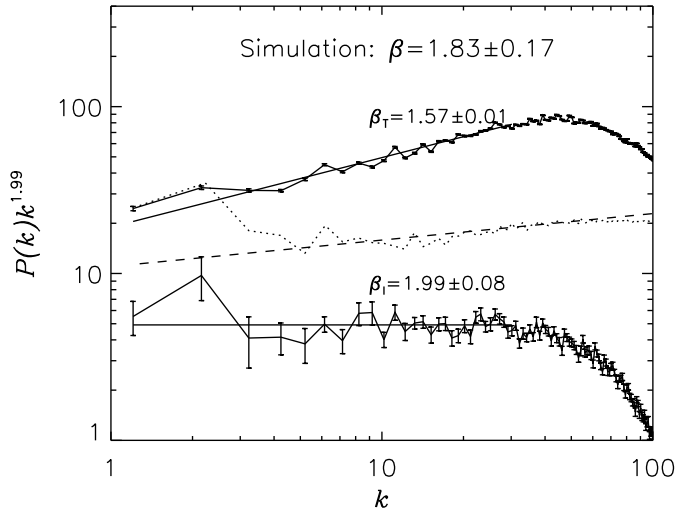}
\caption[]{Compensated power spectrum of integrated intensity (bottom curve) 
and average of the compensated power spectra of single-velocity-channel 
intensity maps (top curve) from the numerical simulation. The dotted plot (middle 
curve) is the 3D power spectrum of velocity and the dashed line shows the 
$\beta=1.83$ slope predicted by the method.}
\label{f1}
\end{figure}
Radioastronomy Observatory (FCRAO) survey of the Perseus molecular cloud 
complex (Ridge et al. 2006), publicly available from the COMPLETE website,
and estimate the power spectrum exponent $\beta=1.81\pm 0.10$. The result
is confirmed by the analysis of the lower resolution map of the same 
region obtained with the AT\&T Bell Laboratories antenna (see Padoan et al. 1999).

\section{Synthetic Spectral Maps}

Synthetic maps of the J=1-0 line of $^{13}$CO are computed with a new non-LTE
radiative transfer code that has been extensively tested against the older 
Monte Carlo code (Juvela 1997). As a model of the density and velocity fields 
in interstellar clouds we use the results of a supersonic simulation 
of HD turbulence with rms Mach number $M=6$. Our purpose is not to simulate
the conditions found in a specific molecular cloud, but rather to test
a general method of estimating the velocity power spectrum. We will study the 
effect of different parameter values elsewhere. Because the method is based on 
an analytical derivation by Lazarian \& Pogosyan (2000) that neglects correlations 
between density and velocity, we expect it to work equally well for a vast range 
of turbulence parameters.

The simulations are carried out with the {\em Enzo} code, developed at the 
Laboratory for Computational Astrophysics by Bryan, Norman and collaborators 
(Norman \& Bryan 1999). {\em Enzo} is a public domain 
Eulerian grid-based code (see http://cosmos.ucsd.edu/enzo/) that adopts the 
Piecewise Parabolic Method (PPM) of Colella \& Woodward (1984). We use an 
isothermal equation of state, periodic boundary conditions, initially uniform 
density and initial random large scale velocity. The turbulence is forced in 
Fourier space only in the wavenumber range $1\le k\le 2$, where $k=1$ corresponds 
to the size of the computational domain that contains $1,024^3$ computational 
zones (for details see Kritsuk et al. 2006). 

\begin{figure}[ht]
\centering
\epsfxsize=9cm \epsfbox{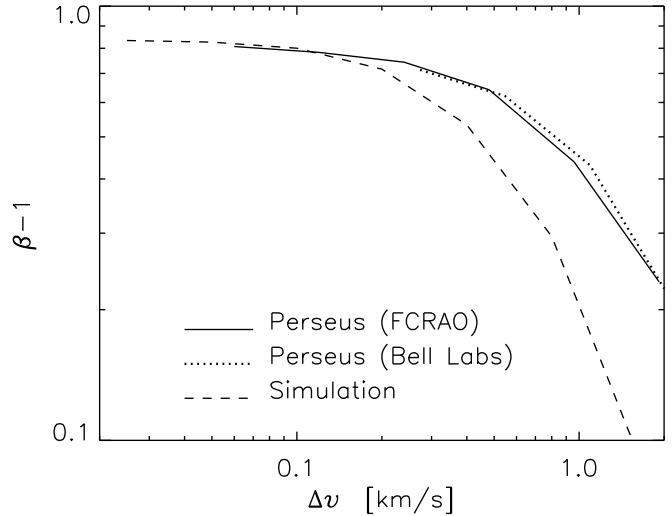}
\caption[]{Convergence plot for the exponent of the velocity power spectrum
as a function of the channel width.}
\label{f2}
\end{figure}

The radiative transfer calculations assume a box size of 5~pc, a mean density 
of $10^3$~cm$^{-3}$, a mean kinetic temperature of 10~K, an rms Mach number 
$M=6$ (consistent with the HD simulation) and a uniform abundance of $^{13}$CO 
molecules. These values were chosen as a generic reference model, not tailored 
to the Perseus molecular cloud complex. The original density and velocity 
data-cubes are re-sampled to a resolution of $256^3$ zones. This has three 
advantages: i) It leaves density and velocity fields with power spectra that 
are power laws almost up to the new Nyquist frequency; ii) It speeds up enormously 
the radiative transfer calculations; iii) It generates a map of synthetic 
spectra with a range of scales comparable to that of the rebinned FCRAO map 
of Perseus (see below).

The result of the radiative transfer calculations is a map of spectral profiles 
of the line intensity, $T(v,{\bf x})$, over 280 velocity channels, $v$, and 
across all positions, ${\bf x}$, within the map of $256^2$ positions. The sum 
of the line intensity of all velocity channels, multiplied by the channel width, 
$\Delta v=0.025$~km/s, gives the integrated intensity at each position, 
$I({\bf x})=\Sigma_v T(v,{\bf x})\,\Delta v$. The map $I({\bf x})$ may be used 
to derive a rough estimate of the column density. We do not address the issue 
of converting integrated intensities into accurate estimates of column density, 
because we derive the velocity power spectrum directly from the line intensity 
data.

\section{The Method}

We call $P_{\rm I}(k)$ the two-dimensional spatial power spectrum of the 
integrated intensity map, $I({\bf x})$, and look for a power law fit such 
that, $P_{\rm I}(k)\propto k^{-\beta_{\rm I}}$. The integrated map is the 
sum of all the single-velocity-channel maps. We then call $P_{\rm T}(v,k)$ 
the two-dimensional spatial power spectrum of the individual channel 
maps, $T(v,{\bf x})$, where we have left the velocity dependence to 
stress that there are as many of these maps and power spectra as there 
are velocity channels. We look for a power law fit to the average of 
these power spectra as well, such that 
$P_{\rm T}(k)=\langle P_{\rm T}(v,k)\rangle_v \propto k^{-\beta_{\rm T}}$. 
Finally, we assume the three dimensional spatial power spectrum of the
\begin{figure}[ht]
\centering
\epsfxsize=9cm \epsfbox{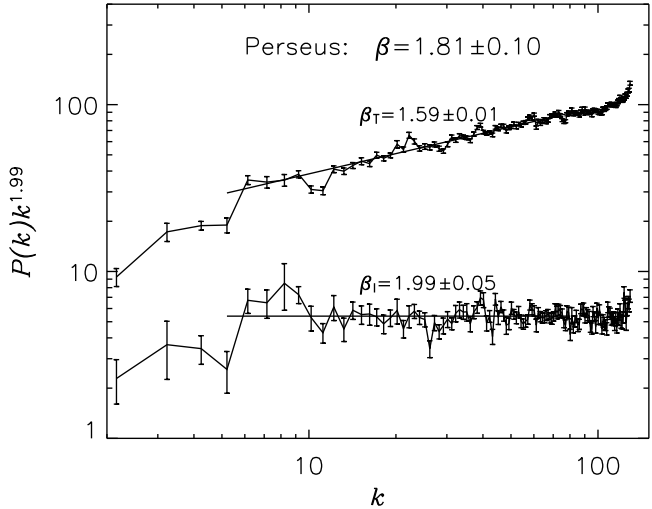}
\caption[]{Same as in Figure~\ref{f1}, but for the FCRAO survey
of Perseus (Ridge et al. 2006).}
\label{f3}
\end{figure}
velocity field is also a power law (a reasonable assumption for the cloud 
turbulence, given the huge Reynolds number), $P_{\rm v}(k)\propto k^{-\beta}$. 

The result of Lazarian \& Pogosyan (2000) can be expressed as 
$\beta=1+2(\beta_{\rm I}-\beta_{\rm T})$, if $\beta_{\rm I}<2$, and 
$\beta=1+2(2-\beta_{\rm T})$, if $\beta_{\rm I}>2$. In this Letter
we test their result only for the case of $\beta_{\rm I}<2$. The 
single-velocity-channel power spectra are shallower than the integrated 
intensity power spectrum, $\beta_{\rm T}<\beta_{\rm I}$, because they 
contain small-scale structure originating from both the density and 
the velocity fields. It is this difference that makes it possible to
derive the velocity power spectrum exponent, based on the above formula.

In this Letter we refer to power spectra as the {\it total} power within 
shells of wavenumber between $k$ and $k+dk$. In Lazarian \& Pogosyan (2000), 
the power spectra are defined as the {\it average} power within shells 
of wavenumber between $k$ and $k+dk$. Furthermore, we do not absorb the 
negative sign in the definition of $\beta_I$ as they do. Our exponents 
are related to their $n$ and $m$ exponents in the following way: 
$\beta_I=1-n$, $\beta=1+m$. In our convention, the velocity power 
spectrum corresponds to the usual turbulent energy spectrum, for example 
$k^{-5/3}$ for the Kolmogorov case. 

Fig.~\ref{f1} shows $P_{\rm I}(k)$ (bottom plot) and $P_{\rm T}(k)$ 
(top plot) for the simulated data-cube ``observed'' along the $x$ direction.
The error bars are one standard deviation above and below the mean values 
and depend on the statistical sample size (number of Fourier modes inside 
each wavenumber shell), so they decrease with increasing wavenumber. The 
linear least square fits, plotted as solid lines, are computed in the 
wavenumber range $1\le k \le 30$. Based on the estimated exponents of 
these two power spectra, we obtain 
$\beta=1+2(\beta_{\rm I}-\beta_{\rm T})=1.83\pm 0.17$. Values of $\beta$ 
from synthetic maps of the data-cube observed in different directions fall 
within the estimated 1-$\sigma$ uncertainty. The value of the 
velocity power spectrum exponent, computed directly from the original 
three dimensional velocity field, is $\beta=1.8\pm 0.1$ (dotted plot in Fig.~\ref{f1}), 
showing that the method retrieves the correct exponent. 

In order to verify the dependence of $\beta$ on the velocity channel 
width, we have applied the method to synthetic maps at different 
velocity resolutions. The result, plotted in Fig.~\ref{f2}, shows 
that the value of $\beta$ is completely converged only for a channel width of 
the order of the thermal width. This was to be expected, because all velocity 
fluctuations above the thermal width may in principle affect the velocity power 
spectra of the single-velocity-channel maps. 

This method is based on an analytical derivation by Lazarian \& Pogosyan (2000)
that neglects the correlations of density and velocity in turbulent flows.
We interpret this test as a confirmation of the validity of their method
in the case of $\beta_{\rm I}<2$, rather than as an empirical calibration
of the value of $\beta$ based on $\beta_{\rm I}$ and $\beta_{\rm T}$. The
uncertainty of the method, when applied to observational data, is then
determined by the error bars of the observational power spectra, 
independently of the uncertainty of our numerical test. The final 
error bar is dominated by the uncertainty in the power spectrum exponent 
of the integrated intensity, because the uncertainty in the 
single-velocity-channel spectrum is reduced by averaging the power 
spectra of many velocity channels.

\section{The Power Spectrum of Perseus}

We have applied the method to the J=1-0 $^{13}$CO survey of the Perseus 
molecular cloud complex carried out with the FCRAO 14~m antenna by 
Ridge et al. (2006). The grid spacing of the survey is 23'', and the beam
size 46''. The velocity-channel size is 0.06~km/s. The power spectra we
compute are corrected for the effect of beam and noise, by simply dividing
by the power spectrum of a gaussian beam, and subtracting the power spectrum
of the noise. Spatial correlations in the noise arising from the 
``On-the-Fly'' mapping mode are neglected.  At
the largest wavenumbers, the power spectra are sensitive to the noise 
subtraction, and realistic error bars accounting for that would make
such wavenumbers essentially useless for estimating the power spectra.
We therefore prefer to regrid the map to a resolution of 92'', which
has the advantage of increasing the signal-to-noise by a factor of 4.

The power spectra are shown in Fig.~\ref{f3}. The least square fits 
are computed in the range $5\le k \le 80$, and yield $\beta=1.81\pm 0.10$. 
Although this range is less extended than that used with the synthetic 
data, it includes larger wavenumbers (between $k=30$ and $k=80$) than 
the synthetic fit (the synthetic power spectra are still affected by 
numerical dissipation at large wavenumbers even after rebinning from 
$1,024^3$ to $256^3$). This reduces the final uncertainty, because 
the statistical sample size is much larger at larger wavenumbers 
($\propto k\,dk$ in two dimensions). As a result, the 1-$\sigma$ 
uncertainty of $\beta$ is 9\% for the synthetic data, and 5\% 
for the observations. 

The value of the slope of the projected density power spectrum, 
$\beta_{\rm I}=1.99\pm 0.05$ is similar to values previously found 
in different regions observed in HI (e.g. Green 1993; Stanimirovich 
et al. 1999) and CO (e.g. Bensch, Stutzki, Ossenkopf 2001; Padoan et al. 2004a). 
Notice that in those previous works, with the exception of
Padoan et al. (2004a), the power spectrum is not integrated over wavenumber
shells, so its slope is equivalent to $\beta_{\rm I}+1$. Furthermore, the power 
spectrum of the $^{13}$CO integrated intensity is slightly steeper than the gas 
density power spectrum due to radiative transfer effects. Accounting for such 
effects, the slope of the projected gas density power spectrum in the Perseus 
region was estimated to be consistent with that of super-Alf\'{e}nic 
turbulence simulations (Padoan et al. 2004a).

The value of $\beta$ estimated for the Perseus region as a function 
of the velocity resolution is shown in Fig.~\ref{f2} (solid line).
The channel width of 0.06~km/s is close to the thermal 
line width and the value of $\beta$ seems to be almost converged,
at least within its 1-$\sigma$ uncertainty. As an independent 
test, we have applied the method also to the AT\&T Bell Laboratories 
map of the same region (see Padoan et al. 1999). The Bell Laboratories
7~m antenna has a beam twice the size of the FCRAO 14~m antenna. We did 
not regrid this map to a lower resolution, so the map resolution and 
size are comparable to those of the FCRAO map, but its velocity resolution 
is only 0.273~km/s, so it is 
not expected to yield a converged value of $\beta$. The values of $\beta$ 
for the Bell Laboratories map is shown in Fig.~\ref{f2} as a dotted line, 
showing very good agreement with the FCRAO result, well within the estimated 
1-$\sigma$ uncertainty.

\section{Discussion and Conclusions}

This result has very interesting implications. First, numerical simulations giving
power spectra with slope significantly larger than $\beta=1.8\pm 0.1$ may be 
ruled out as correct descriptions of molecular cloud turbulence (at least for 
the Perseus region). Burgers exponent, $\beta=2.0$, is 2$\sigma$ larger than 
the Perseus exponent (assuming this is converged as a function of velocity 
resolution). The slope of the power spectra of the SPH simulations in 
Ballesteros-Paredes et al. (2006), $\beta\approx 2.7$ in the case of a turbulence 
rms Mach number $\approx 6$, is 9$\sigma$ larger than the present estimate, 
and their grid based simulations have $\beta\approx 2.2$, 4$\sigma$ too large. 
Second, we can now derive the absolute values of the velocity structure function 
exponents in Perseus. Padoan et al. (2004b) have determined numerically the
relative values of those exponents, so the knowledge of one of them, for example
the second order exponent given by $\beta-1$, allows to determine the absolute
values of all the others. 

A different method of estimating the scaling of the turbulence from molecular
clouds surveys was developed by Brunt \& Heyer (2002a,b), based on the principle 
component analysis (PCA). They analyzed 23 molecular clouds in the outer
Galaxy and estimated a value $\beta=2.2 \pm 0.3$, if the structure 
function exponents of order $p$ are assumed to scale linearly as $p/3$. This value 
is significantly larger than the one we estimate in Perseus. However, the PCA 
method is dependent on a calibration with numerical simulations. Based on such
a calibration, it appears that the PCA method estimates the exponent of the velocity 
structure functions of order $p=0.5$ or lower (Brunt et al. 2003). Taking this into 
consideration, the result of Brunt \& Heyer (2002b) would be roughly consistent with 
ours, if a very intermittent scaling of the structure function is assumed, consistent 
with numerical simulations of supersonic turbulence. 

We have shown in this Letter that the method for estimating the velocity power 
spectrum slope proposed by Lazarian and Pogosyan (2000) works well in the case
of $\beta_{\rm I}<2$, and for a velocity resolution not much larger than the 
thermal line width. However, regions with lower turbulence Mach number than
Perseus, for example the Taurus region, may have steeper density power spectra, 
and hence $\beta_{\rm I}>2$. Such regions will be studied in a future work, where
the method will be tested also with numerical simulations generating synthetic data
with $\beta_{\rm I}>2$.

\acknowledgements

P.P., A.K., and M.L.N. were partially supported by the NASA ATP grant
NNG056601G, the NSF grants AST-0507768 and AST-0607675 and the NRAC 
allocation MCA098020S. We utilized computing resources provided by the 
San Diego Supercomputer Center and by the National Center for Supercomputing 
Applications. M.J. was supported by the Academy of Finland Grants no. 206049 
and 107701.

\bibliographystyle{apj}

\begin{thebibliography}

\bibitem[Bensch et al.(2001)]{2001A&A...366..636B} Bensch, F., Stutzki, J., 
\& Ossenkopf, V.\ 2001, \aap, 366, 636 

\bibitem[Boldyrev(2002)]{Boldyrev2002}
Boldyrev, S. 2002, \apj, 569, 841

\bibitem[Brunt \& Heyer(2002a)]{Brunt+Heyer2002calibration}
Brunt, C.~M. \& Heyer, M.~H. 2002a, \apj, 566, 276
 
\bibitem[Brunt \& Heyer(2002b)]{Brunt+Heyer2002results}
Brunt, C.~M. \& Heyer, M.~H. 2002b, \apj, 566, 289

\bibitem[Brunt et al.(2003)]{2003ApJ...595..824B} Brunt, C.~M., Heyer, 
M.~H., V{\'a}zquez-Semadeni, E., \& Pichardo, B.\ 2003, \apj, 595, 824 

\bibitem[Colella \& Woodward(1984)]{1984JCoPh..54..174C} Colella, P., \& 
Woodward, P.~R.\ 1984, Journal of Computational Physics, 54, 174 

\bibitem[Dobler et al.(2003)]{2003PhRvE..68b6304D} Dobler, W., Haugen, 
N.~E., Yousef, T.~A., \& Brandenburg, A.\ 2003, \pre, 68, 026304 

\bibitem[Esquivel et al.(2003)]{2003MNRAS.342..325E} Esquivel, A., 
Lazarian, A., Pogosyan, D., \& Cho, J.\ 2003, \mnras, 342, 325 

\bibitem[Falkovich(1994)]{1994PhFl....6.1411F} Falkovich, G.\ 1994, Physics 
of Fluids, 6, 1411 

\bibitem[Frisch \& Bec(2001)]{2001ntt..conf..341F} Frisch, U., \& Bec, J.\
2001, New trends in turbulence.~Editors: M.~Lesieur, A.~Yaglom, F.~David,
Les Houches Summer School, vol.~74, p.341, 341

\bibitem[Green(1993)]{1993MNRAS.262..327G} Green, D.~A.\ 1993, \mnras, 262, 
327

\bibitem[Haugen \& Brandenburg(2004)]{2004PhRvE..70b6405H} Haugen, N.~E., 
\& Brandenburg, A.\ 2004, \pre, 70, 026405 

\bibitem[Juvela(1997)]{1997A&A...322..943J} Juvela, M.\ 1997, \aap, 322, 
943 

\bibitem[Klein et al.(2006)]{} Klein, R.~I., Inutsuka, S., Padoan, P., 
\& Tomisaka, K.\ 2006, in Protostars and Planets V, eds. B. Reipurth et al.

\bibitem[Kritsuk et al.(2006)]{}Kritsuk, A. G., Wagner, R., Norman, M. L., 
\& Padoan, P.\ 2006, in ASP Conf. Ser., Calspace-IGPP Conf. on Numerical 
Modeling of Space Plasma Flows (San Francisco: ASP), in press

\bibitem[Lazarian \& Pogosyan(2000)]{2000ApJ...537..720L} Lazarian, A., \& 
Pogosyan, D.\ 2000, \apj, 537, 720 

\bibitem[Mizuno et al.(1995)]{1995ApJ...445L.161M} Mizuno, A., Onishi, T., 
Yonekura, Y., Nagahama, T., Ogawa, H., \& Fukui, Y.\ 1995, \apjl, 445, L161 

\bibitem[Nordlund \& Padoan(2003)]{2003LNP...614..271N} Nordlund, {\AA}., 
\& Padoan, P.\ 2003, LNP Vol.~614: Turbulence and Magnetic Fields in 
Astrophysics, 614, 271 

\bibitem[Norman \& Bryan(1999)]{1999numa.conf...19N} Norman, M.~L., \& 
Bryan, G.~L.\ 1999, ASSL Vol.~240: Numerical Astrophysics, 19 

\bibitem[Padoan et al.(1999)]{1999ApJ...525..318P} Padoan, P., Bally, J., 
Billawala, Y., Juvela, M., \& Nordlund, {\AA}.\ 1999, \apj, 525, 318 

\bibitem[Padoan et al.(2004)]{2004ApJ...604L..49P} Padoan, P., Jimenez, R., 
Juvela, M., \& Nordlund, {\AA}.\ 2004a, \apjl, 604, L49 

\bibitem[Padoan et al.(2004)]{2004PhRvL..92s1102P} Padoan, P., Jimenez, R., 
Nordlund, {\AA}., \& Boldyrev, S.\ 2004b, Physical Review Letters, 92, 
191102

\bibitem[Ridge et al.(2006)]{2006AJ....131.2921R} Ridge, N.~A., et al.\ 
2006, \aj, 131, 2921 

\bibitem[Stanimirovic et al.(1999)]{1999MNRAS.302..417S} Stanimirovic, S., 
Staveley-Smith, L., Dickey, J.~M., Sault, R.~J., \& Snowden, S.~L.\ 1999, 
\mnras, 302, 417 

\end{thebibliography}

\end{document}